\documentclass[12pt]{article}
\usepackage{graphicx}
\usepackage{amsmath,amssymb}
\usepackage{ulem}
\newcommand{\pa}{\partial}
\newcommand{\ts}{\tilde{s}}
\newcommand{\tx}{\tilde{x}}
\newcommand{\tu}{\tilde{u}}

\newcommand{\tp}{\tilde{p}}
\newcommand{\tpa}{\tilde{\partial}}

\begin{document}
\begin{titlepage}
\begin{flushright}
TIT/HEP-672 \\
March,  2019
\end{flushright}
\vspace{0.5cm}
\begin{center}
{\Large \bf
Quantum Seiberg-Witten curve and Universality \\
in
Argyres-Douglas theories 
}
\lineskip.75em
\vskip 2.5cm
{\large  Katsushi Ito, Saki Koizumi and Takafumi Okubo }
\vskip 2.5em
{\normalsize\it Department of Physics,\\
Tokyo Institute of Technology\\
Tokyo, 152-8551, Japan}
\vskip 3.0em
\end{center}
\begin{abstract}
We study the quantum Seiberg-Witten (SW) curves for $(A_1, G)$-type Argyres-Douglas (AD) theory  by taking the scaling limit of the quantum SW curve of ${\cal N}=2$ gauge theory with gauge group $G$.
For $G=A_r$, the quantum SW curve of the AD theory is consistent with the scaling limit of the curve of the gauge theory. 
For $G=D_r$,  we need the quantum correction to the SW curve of the AD theory, which depends on the quantization condition of the original SW curve. 
We also study the universality of the quantum SW curves for $(A_1, A_3)$ and  $(A_1, D_4)$-type AD theories.
\end{abstract}
\end{titlepage}
\baselineskip=0.7cm

\section{Introduction}
The low-energy effective action of four-dimensional ${\cal N}=2$ supersymmetric gauge theory is determined by the prepotential,  which is a function of the
period integrals of the Seiberg-Witten (SW) differential  on a Riemann surface
called the SW curve
\cite{Seiberg:1994rs}.
The prepotential deformed in the $\Omega$-background 
has a rich mathematical structure \cite{Nekrasov:2002qd}.  
In particular, in the Nekrasov-Shatashvili (NS) limit of the $\Omega$-background,
the deformed periods are characterized by the Bohr-Sommerfeld type quantization condition \cite{Nekrasov:2009rc}.
This condition leads to the idea of quantization of the SW curve \cite{Mironov:2009uv}.
The quantum SW curve is a differential equation defined by the symplectic structure on the curve. 
It can be also regarded as  Baxter's T-Q relation of an integrable system \cite{Mironov:2009uv,Mironov:2009dv,Popolitov:2010bz,Zenkevich:2011zx}.
The quantum SW periods for various ${\cal N}=2$ gauge theories have been studied
with the help of topological strings \cite{Huang:2009md,Huang:2012kn,Krefl:2014nfa}, CFT \cite{Maruyoshi:2010iu,Piatek:2014lma,Poghossian:2016rzb} and  the exact WKB method \cite{Basar:2015xna,Ashok:2016yxz,Basar:2017hpr}.

The quantum SW curve provides also a useful tool to study the 
$\Omega$-deformed theory in the strong coupling 
region
\cite{He:2010xa,Ito:2017iba}. 
In particular,  the Argyres-Douglas (AD) theories realized at the RG fixed points in the Coulomb branch of the moduli space, are a recent interesting subject of study \cite{Argyres:1995jj, Argyres:1995xn}. 
The SW curves of the AD theories are obtained by degeneration of the SW curves of the corresponding gauge theories, which are classified in \cite{Eguchi:1996vu,Cecotti:2010fi,Xie:2012hs}.
For example, $(A_1,G)$-type AD theories for a simply-laced Lie algebra $G$ are obtained by degeneration of the SW theory with gauge group $G$.

The purpose of this paper is to study the quantum SW curve for the AD theories,
 which are obtained by the scaling limit of the gauge theories.
For the AD theories obtained by the scaling limit of ${\cal N}=2$ $SU(2)$
SQCD with $N_f=1,2,3$ hypermultiplets \cite{Argyres:1995xn}, the deformed SW periods
have been calculated in \cite{Ito:2018hwp}. 
It has been shown that the quantization conditions of the curves depend on the flavor symmetry and the quantum SW curves take different forms.
It has been known that there exists universality in AD 
theories, where
different UV gauge theories correspond to the same AD theory.
For example, the $SU(3)$ $N_f=0$ theory and the $SU(2)$ $N_f=1$ theory  
lead to the ($A_1,A_2$)-type AD theory, the $SU(4)$  theory and $SU(2)$ $N_f=2$ theory to the ($A_1, A_3$) type
, $SO(8)$  theory and $SU(2)$ $N_f=3$ theory to the ($A_1,D_4$) type.  
Since the quantum SW curves of $A_r$ theory and SQCD are based on the different quantization conditions, it is interesting to check the universality of AD theories at the quantum level.
Moreover, in the $D_r$ theories, there are various ways to quantize the SW curve, which provide different quantum corrections to the SW periods in the scaling limit.
It would be interesting to study the relation among the quantization conditions for the
AD theories.

In this paper, we study the quantum SW curve for the AD theories of type ($A_1,G$) for $G=A_r$ and $D_r$ by taking the scaling limit of ${\cal N}=2$  super Yang-Mills theory
with gauge group $G$. 
Based on the quantum SW curves for the AD theories, 
we will discuss that the universality of $(A_1, A_3)$ and $(A_1,D_4)$ type AD theories. 

This paper is organized as follows:
in section 2, we study the scaling limit of the quantum SW curve for the $A_r$ gauge theory. In particular, we discuss the universality of 
$(A_1, A_3)$ type AD theory.
In section 3, we study the quantum SW curve for $D_r$-type AD theories.
We discuss the relation to the $(A_1,D_4)$-type AD theory obtained from the scaling limit of $SU(2)$ $N_f=3$ theory.
Section 4 is for summary and discussion.

\section{Quantum SW curve of $A_r$-type AD theories}
In this section, we study  the quantum SW curve for ${\cal N}=2$ super Yang-Mills theory with $A_r$-type gauge group and its scaling limit around the superconformal point \cite{Argyres:1995jj,Eguchi:1996vu}.
The SW curve for $A_r$ type gauge group  is defined by \cite{Klemm:1994qs,Argyres:1994xh,Martinec:1995by,Ito:1999cc}
\begin{align}
\frac{\Lambda^{r+1}}{2}
\left(z+\frac{1}{z}\right)=W(x,u_i), \label{11.22.10}
\end{align}
where
\begin{align}
W(x,u_i)&=x^{r+1}-u_2 x^{r-1}-\cdots -u_{r+1}.
\end{align}
Here  
$u_i$'s are the Coulomb moduli parameters and $\Lambda$ is the QCD scale parameter.
The SW differential $\lambda_{SW}$ is defined by
\begin{align}
\lambda_{SW}=x\frac{dz}{z}.
\end{align}
The SW periods are given by the following integrals 
\begin{align}
\Pi^{(0)}=(a^{(0)}_I,a^{(0)}_{DI})=
\left(
\oint_{\alpha_I} \lambda_{SW},\oint_{\beta_I} \lambda_{SW}
\right),\qquad I=1,\ldots, r
\label{2019.1.4}
\end{align}
where $\alpha_I$ and $\beta_I$ are 1-cycles on the curve with
canonical intersection numbers.
We refer the superscript $(0)$ to represent the undeformed or classical SW periods.

There are singularities on the moduli space where some mutually nonlocal
BPS particles become massless.
These singular points are called the superconformal points.
For $A_r$ type SW curve (\ref{11.22.10}), they are given by \cite{Ito:1999cc}
\begin{align}
    u_2&=\cdots=u_r=0,\quad u_{r+1}=\pm \Lambda^{r+1}.
\end{align}
We consider the scaling limit around the superconformal point $u_{r+1}=-\Lambda^{r+1}$.
We introduce the scaling parameters by 
\begin{align}
z&=e^{\delta \xi}, \qquad
x=\delta^{\frac{2}{r+1}}\Lambda\tilde{x}, \nonumber\\
u_i&=\delta^{\frac{i}{r+1}}\Lambda^i\tilde{u}_i, \qquad i=2,\ldots, r,\nonumber \\
u_{r+1}&=-\Lambda^{r+1}+\delta^{\frac{2(r+1)}{ r+1}}\Lambda^{r+1}\tilde{u}_{r+1} 
\label{eq:scalingar}
\end{align}
and take the limit $\delta\rightarrow 0$ \cite{Ito:1999cc,Grassi:2018bci}.
In the scaling limit, 
the SW curve (\ref{11.22.10})  becomes
\begin{align}
\xi^2&=2 W(\tilde{x},\tilde{u}_i). \label{eq:swc_ad_ar}
\end{align}
The SW differential scales as $\lambda_{SW}=\delta^{r+3\over r+1}\Lambda\tilde{\lambda}_{SW}$, where
\begin{align}
\tilde{\lambda}_{SW}&= \tilde{x}d\xi. \label{eq:swd_ad_ar}
\end{align}
The curve (\ref{eq:swc_ad_ar}) describes the AD theory of $(A_1,A_r)$-type, where the parameter $\tilde{u}_j$ has 
scaling dimension $2j/(r+3)$ ($j=2,\ldots, r+1$).
The SW periods for the curve are defined by
\begin{align}
    \tilde{\Pi}^{(0)}&=(\tilde{a}_I^{(0)},\tilde{a}_{DI}^{(0)})=
    \left(
\oint_{\tilde{\alpha}_I} \tilde{\lambda}_{SW},\oint_{\tilde{\beta}_I} \tilde{\lambda}_{SW}
\right),\qquad I=1,\ldots, \left[{r\over2}\right].
\end{align}

Next, we study the quantum corrections to the SW period $\Pi^{(0)}$.
For the SW curve (\ref{11.22.10}), the symplectic form  is
defined by $d\lambda_{SW}=dx\wedge d\log z$.
We therefore quantize 
the curve by replacing 
${\rm log}z\to -i\hbar\partial_x$.
We then obtain the quantum SW curve \cite{Grassi:2018bci}
\begin{align}
\left[\frac{\Lambda^{r+1}}{2}
\left(e^{-i\hbar\partial_x}+e^{i\hbar\partial_x}\right)
-W(x,u_i)\right]\Psi(x)=0.
\label{11.30.10}
\end{align}
Note that in \cite{Mironov:2009dv} the quantum SW curve has been obtained from the quantization $x\rightarrow -i\hbar \partial_{\log z}$.
But both the quantum SW periods are shown to be the same.
The WKB solution to this equation is of the form:
\begin{align}
\Psi(x)&=\exp\left(
{i\over \hbar}\int^x dx P(x)
\right), \quad P(x)=\sum_{n=0}^{\infty} \hbar^n p_n(x).\label{11.30.11}
\end{align}
Substituting this solution into (\ref{11.30.10}), we can  determine $p_n$
recursively.
We can show that 
$p_0dx$ is nothing but the SW differential $\lambda_{SW}$ up to total derivatives.
For odd $n$, $p_{n}(x)$ takes the form of total derivatives.
The first three terms of $p_{2n}(x)$ are given by 
\begin{align}
p_0&=\log (B-\sqrt{B^2-1}) \label{eq:p0}, \\
p_2
&={\partial_x^2 B \over 24 (B^2-1)^{5\over2}}+d(*), \label{eq:p2}
\\
p_4&=-{B (23+12 B^2) \over 1920 (B^2-1)^{7\over2}} (\partial_x^2 B)^2
+{(7+8 B^2) \over 5760 (B^2-1)^{5\over2}} \partial_x^4 B+d(*),
\label{eq:p4}
\end{align}
where $B:=W/\Lambda^{r+1}$.
We define the quantum SW periods by
\begin{align}
    \Pi(\hbar)&=\left(\int_{\alpha_I}P(x)dx, \int_{\beta_I}P(x)dx\right)=\sum_{n=0}^{\infty}\hbar^{2n} \Pi^{(2n)}
\end{align}
where
\begin{align}
    \Pi^{(2n)}&=\left(\int_{\alpha_I}p_{2n}(x)dx, \int_{\beta_I}p_{2n}(x)dx\right).
\end{align}

Now we consider the quantum SW curve for the AD theory.
The symplectic structure is defined by $d\tilde{\lambda}_{SW}=d\tilde{x}\wedge d\xi$.
Replacing $\xi\to -i\hbar\partial_{\tilde{x}}$,
the curve (\ref{eq:swc_ad_ar}) becomes the quantum SW curve for the $A_r$-type AD theory
\begin{align}
\left[-\hbar^2\frac{\partial^2}{\partial \tilde{x}^2}-2W(\tilde{x},\tilde{u}_i)\right]\tilde{\Psi}(\tilde{x})=0.
\label{11.22.5}
\end{align}
This is the Schr\"odinger type equation.
The WKB solution to  (\ref{11.22.5}) 
is given by
\begin{align}
\tilde{\Psi}(\tilde{x})&=\exp\left(
{i\over \hbar}\int^{\tilde{x}} dx \tilde{P}(x)
\right), \quad \tilde{P}(\tilde{x})=\sum_{n=0}^{\infty} \hbar^n \tilde{p}_n(\tilde{x}).\label{eq:wkb_ar}
\end{align}
Here $\tilde{p}_n(\tilde{x})$ are obtained by the WKB expansion of the Schr\"odinger 
equation (see \cite{Ito:2017iba} for example). 
The first few coefficients are 
\begin{align}
\tilde{p}_0&=-\sqrt{2}\sqrt{W},
\nonumber\\
\tilde{p}_2&=\frac{1}{48\sqrt{2}}\frac{\partial_{\tilde{x}}^2W}{W^{3/2}},
\nonumber\\
\tilde{p}_4&={7\over 3072\sqrt{2}}{(\pa_{\tilde{x}}^2 W)^2 \over W^{7\over2}}
-{1\over 1536\sqrt{2}}{\pa_{\tilde{x}}^4 W\over W^{5\over2}}
, \label{eq:pn_ar_ad}
\end{align}
up to total derivatives.
The quantum SW periods of the AD theory are also defined by
\begin{align}
    \tilde{\Pi}(\hbar)&=\sum_{n=0}^{\infty}\hbar^{2n} \tilde{\Pi}^{(2n)},\quad
    \tilde{\Pi}^{(2n)}=\left(\int_{\alpha_I}\tilde{p}_{2n}(\tilde{x})d\tilde{x},
    \int_{\beta_I}\tilde{p}_{2n}(\tilde{x})d\tilde{x}\right).
\end{align}
We compare these corrections with the scaling limit of the quantum SW curve (\ref{11.30.10}) of $A_r$-type gauge theory.
Taking the scaling limit of (\ref{eq:p0})-(\ref{eq:p4}),
we find that
\begin{align}
p_{2n}(x)dx&=\delta^{{(1-2n)(r+3)\over r+1}}\Lambda^{-n}\tilde{p}_{2n}(\tilde{x})d\tilde{x}+\cdots,    
\end{align}
up to total derivatives.

We next study the relations among the quantum SW periods in the $A_r$ type AD theory.
It is convenient to 
find the relation between $\tilde{p}_{2n}$ and $\tilde{p}_0$
\begin{align}
\tilde{p}_{2n}=\tilde{\cal O}_{2n}\tilde{p}_0,
\end{align}
where $\tilde{\cal O}_{2n}$ is a differential operator with respect to $\tilde{u}_i$.
This expression is useful to evaluate higher order corrections because the integral of $\tilde{p}_{2n}$ contains superficial divergence at the endpoint of the integration region.
Such operators have been calculated for $SU(N_c)$ SQCD \cite{Mironov:2009uv,Mironov:2009dv,Popolitov:2010bz,He:2010xa,Zenkevich:2011zx,Huang:2012kn,Basar:2015xna,Basar:2017hpr,Ito:2017iba} and the AD theories associated with $SU(2)$ SQCDs \cite{Ito:2018hwp}. 
Here we will propose a general procedure to find the differential 
operators for $A_r$ and $D_r$ type AD theories.

The derivative of the superpotential $W$ with respect to the moduli parameters $\tilde{u}_i$
 is
 \begin{align}
 \partial_{\tilde{u}_i}W=-x^{r+1-i}.
 \label{eq:sup_del1}
 \end{align}
 This is used to derive the following formula:
\begin{align}
\pa_{\tu_{j_1}}\pa_{\tu_{j_2}} \cdots \pa_{\tu_{j_n}}\tilde{W}^{1\over2}&={(-1)^{n-1} (2n-3)!! \over 2^n}
 {\pa_{\tu_{j_1}}\tilde{W}\cdots \pa_{\tu_{j_n}}\tilde{W} \over \tilde{W}^{n-{1\over2}}}.
 \label{eq:sup_del2}
\end{align}
We expand the $\tilde{x}$ derivatives of $W$ in terms of the products of $\partial_{\tilde{u}_i}W$. 
 Using (\ref{eq:sup_del1}), (\ref{eq:sup_del2}) and (\ref{eq:pn_ar_ad}), we get
 \begin{align}
 \tilde{p}_2
 &=-{1\over 24}
 \left(
  -(r+1) r \pa_{\tu_2} \pa_{\tu_{r+1}}
 + \sum_{j=2}^{r-1} (r+1-j)(r-j) \tu_j \pa_{\tu_{j+2}} \pa_{\tu_{r+1}} 
 \right) \tilde{p}_0,
 \end{align}
\begin{align}
\tilde{p}_4
=&
\left(\frac{7}{5760}
 \Bigl\{
 (r+1)^2 r^2 \pa_{\tu_2}^2 
 -2(r+1)r \sum_{j=2}^{r-2}(r+1-j) (r-j)\tu_j \pa_{\tu_2}\pa_{\tu_{j+2}}
 \right.
 \nonumber\\
&+
 \sum_{j=2}^{r-2}  \sum_{k=2}^{r-2}(r+1-k) (r-k) (r+1-j) (r-j) \tu_j \tu_k
 \pa_{\tu_{j+2}}\pa_{\tu_{k+2}}
 \Bigr\}
 \pa_{\tu_{r+1}}^2
 \notag \\
&\left.+\frac{1}{1152}
 \Big\{ -{(r+1)! \over (r-3)!}
\pa_{\tu_4}
+\sum_{j=2}^{r-3}
{(r+1-j)!\over (r-3-j)!}
\tu_j  \pa_{\tu_{j+4}}
\Bigr\}  \pa_{\tu_{r+1}}^2
\right) \tilde{p}_0.
\end{align}
We can express the above equations by linear combination of $\partial_{\tilde{u_i}}\tilde{p}_0$
by using the Picard-Fuchs equations satisfied by $\tilde{p}_0$ \cite{Ito:1999cc}.

Let us discuss some examples. 
For $r=2$, we have the $(A_1,A_2)$-type AD theory.
$\tilde{\Pi}^{(2)}$ and $\tilde{\Pi}^{(4)}$ satisfy
\begin{align}
    \tilde{\Pi}^{(2)}&={1\over4} \partial_{\tilde{u}_2}
    \partial_{\tilde{u}_3} \tilde{\Pi}^{(0)},
    \nonumber\\
    \tilde{\Pi}^{(4)}&={7\over160} \partial_{\tilde{u}_2}^2
    \partial_{\tilde{u}_3}^2 \tilde{\Pi}^{(0)}.
\end{align}
These corrections agree with those of $SU(2)$ $N_f=1$ AD theory \cite{Ito:2018hwp}, where the SW curve is given by
$y^2=x^3 -2\tilde{m}x-\tilde{u}$. 
The SW differential is given by $y dx$.
The relation between the parameters are $\tilde{u_2}=2\tilde{m}$, $\tilde{u}_3=\tilde{u}$.

For $r=3$, which corresponds to the $(A_1,A_3)$-type AD theory, we obtain
\begin{align}
  \tilde{\Pi}^{(2)}&=\left({1\over2} \partial_{\tilde{u}_2}
    \partial_{\tilde{u}_4}-{1\over12} \tilde{u}_2 \partial_{\tilde{u}_4}^2 \right)\tilde{\Pi}^{(0)},
    \nonumber\\
    \tilde{\Pi}^{(4)}&=
    \left({7\over40} \partial_{\tilde{u}_2}^2
    \partial_{\tilde{u}_4}^2-{7\over120} \tilde{u}_2\partial_{\tilde{u}_2}
    \partial_{\tilde{u}_4}^3+{7\over1440}\tilde{u}_2^2
    \partial_{\tilde{u}_4}^4-{1\over48}\partial_{\tilde{u}_4}^3
    \right)\tilde{\Pi}^{(0)}.   
\end{align}
These equations  agree with those obtained from the $SU(2)$ $N_f=2$ AD theory after some identification of the parameters \cite{Ito:2018hwp}.
The SW curve is
\begin{align}
    y^2&=x^3-2\tilde{m}x^2-\left(\tilde{u}-{4\tilde{m}^2\over3}\right) x
    -{\tilde{C}_2\over4}
\end{align}
where $\tilde{m}$ and $\tilde{u}$ are the coupling and the operator, $\tilde{C}_2$ is the Casimir with degree 2.
These parameters are related to $\tu_i$'s by
\begin{align}
    \tu&=-4\tu_4+{\tu_2^2\over3},\quad \tilde{m}=-\tu_2,\quad\tilde{C}_2=-4\tu_3^2.
\end{align}
The SW differential is $yd\log x$.
We quantize the curve by $y\rightarrow -i\hbar {\pa \over \pa \xi}$ with $x=e^\xi$:
\begin{align}
    \left\{-\hbar^2\frac{\partial^2}{\partial\xi^2}
    -e^{3\xi}+2\tilde{m}e^{2\xi}+\left(\tilde{u}-{4\tilde{m}^2\over3}\right) e^\xi
    +{\tilde{C}_2\over4}\right\} \psi(\xi)=0.
    \label{eq:qc_exp}
\end{align}
Note that two quantum curves (\ref{11.22.5}) and (\ref{eq:qc_exp}) are based on different quantization conditions. Therefore the correspondence between the quantum periods is nontrivial.
For $A_3$ case, it is based on the Schr\"odinger type (\ref{11.30.10}). 
On the other hand, for $SU(2)$ $N_f=2$ case, it is the exponential
type (\ref{eq:qc_exp}), which respects $U(2)$ flavor symmetry manifest.

\section{Quantum SW curve for $D_r$-type AD theory}
In this section, we study the quantum SW curve for $(A_1,D_r)$-type AD theory.

The SW curve and the SW differential for ${\cal N}=2$ theory with $D_r$-type gauge group are given by 
\cite{Brandhuber:1995zp,Martinec:1995by}
\begin{align}
&\frac{1}{2}\left(
z+\frac{G(x)}{z}\right)=C(x)
, \label{eq:swc_dr}
\\
&\lambda_{SW}=xd\log z-\frac{1}{2}xd\log G,
\label{11.3.24}
\end{align}
where 
\begin{align}
C(x)=&x^{2r}+s_1x^{2r-2}+\cdots+s_{r-1}x^2+s_r,\label{11.3.21}
\\
G(x)=&\Lambda^{4r-4}x^4.
\label{11.3.22}
\end{align}
Here
$s_i$ ($i=1,\ldots, r-1$) and $s_r^{1/2}$ are the moduli parameters associated with the Casimir operators of $D_r$. 
Introducing $z=\sqrt{G}w$, we get 
\begin{align}
&\frac{\Lambda^{2r-2}}{2}
\left(w+\frac{1}{w}\right)=W(x,s_i), \label{11.3.25}\\
&\lambda_{SW}=xd{\rm log}w, \label{swdiff_dr}
\end{align}
where $W(x,s_i)$ is defined by 
\begin{align}
W(x,s_i)=\frac{C(x)}{x^2}=x^{2r-2}+s_1x^{2r-4}+\cdots
+s_{r-1}+\frac{s_r}{x^2}.
\end{align}

We study the scaling limit of the curve (\ref{11.3.25}) around a superconformal point in the moduli space.
The superconformal points are given by \cite{Eguchi:1996vu}
\begin{align}
    s_1&=s_2=\cdots=s_{r-2}=s_r=0,\quad s_{r-1}=\pm \Lambda^{2r-2}.
    \label{eq:scfpt_dr}
\end{align}
Then we introduce the scaling parameters by 
\begin{align}
w&=e^{\delta\xi},
\qquad
x=\delta^{\frac{2}{2r-2}}\Lambda\tilde{x},
\nonumber\\
s_i&=\delta^{\frac{4i}{2r-2}}\Lambda^{2i}\tilde{s}_i,\quad
i=1,\cdots,r-2,
\nonumber\\
s_{r-1}&=\Lambda^{2r-2}+\delta^2\Lambda^{2(r-1)}\tilde{s}_{r-1},
\qquad
s_r=\delta^{\frac{4r}{2r-2}}\Lambda^{2r}\tilde{s}_r, \label{eq:scaling_dr}
\end{align}
and take the limit $\delta\rightarrow 0$ \cite{Ito:1999cc}. Here we have chosen the plus sign in (\ref{eq:scfpt_dr}).
Taking the scaling limit, the SW curve (\ref{11.3.25}) and the corresponding SW differential (\ref{swdiff_dr}) become the following:
\begin{align}
\xi^2=&2
W(\tilde{x},\tilde{s}_i),
\label{11.4.40}
\\
\tilde{\lambda}_{SW}&=\tilde{x}d\xi.
\label{eq:swdiff_dr_ad}
\end{align}
This is the SW curve of the $(A_1,D_r)$-type AD theory,
where $\tilde{s}_j$ has the scaling dimension $2j/r$ $(j=1,\ldots, r)$.

We next discuss the quantization of the SW curve of $D_r$-type gauge theory. 
There are some possibilities to represent the SW differential and its symplectic structure for $D_r$-type gauge theory.
The SW curve (\ref{eq:swc_dr}) is regarded as the truncation of parameters of moduli parameters of $SU(2r)$ 
gauge theory with massless $N_f=4$ hypermultiplets. 
In the SW differential (\ref{11.3.24}), the flavor part and the moduli part are separated. 
Therefore when we regard $x$ and $\log z$ as the canonical variables, the quantization respects the $SU(4)$ flavor 
symmetry.
On the other hand, in the curve of the form (\ref{11.3.25}),  
the flavor and the moduli parameters are 
included in the same superpotential $W(x,s_i)$. 
Then the quantization based on the canonical variables $(x,\log w)$ do not respect
the flavor symmetry manifestly.
We will see these different choices of the canonical variables lead to the different quantum corrections to the SW periods.

If we quantize the SW curve (\ref{11.3.25}) by $\log w\rightarrow -i\hbar \partial_x $, we obtain the quantum SW curve 
\begin{align}
    \left[ {\Lambda^{2r-2}\over2} \left( e^{-i\hbar \partial_x}+e^{i\hbar \partial_x} \right)-W(x,s_i) \right]\Psi(x)=0.
    \label{eq:qswc_dr1}
\end{align}
Then the WKB solution is calculated as in (\ref{eq:p0})-(\ref{eq:p4}), where $B$ is defined as $W(x,s_i)/\Lambda^{2r-2}$.
If we quantize the SW curve (\ref{eq:swc_dr}) by
$
\log z\to -i\hbar\partial_x,
$
we get the quantum SW curve:
\begin{align}
\frac{1}{2}
\left[e^{-i\hbar\partial_x}+e^{i\hbar\partial_x/2}G(x)e^{i\hbar\partial_x/2}\right]\Psi(x)
=C(x)\Psi(x).
\label{eq:qsw_dr2}
\end{align}
Here we have chosen the ordering of the operators as in \cite{Zenkevich:2011zx}.
Let us consider the WKB solution (\ref{11.30.11}) to the differential equation (\ref{eq:qsw_dr2}).
We find that $p_n$'s are determined recursively and observe that $p_{2n+1}(x)$ is total derivative.
The first three $p_{2n}$'s are given by 
\begin{align}
p_0&=\log(C-y),\label{1.16.1}
\\
p_2&=-{\partial_x^2C\over 8y}+{C (\partial_x y)^2\over 8 y^3},\label{1.16.2}
\\
p_4&=-{1\over 384 y^7}
\left\{
12 C (C')^2 y^2 (y')^2-60 C^2 C' y (y')^3
+12 y^3 C' (y')^3+45 C^3 (y')^4-15 C y^2 (y')^2
\right.
\notag\\
&-12 y^4 C' C''
+24 C C' C'' y^3 y'+42 C^2 C''y^2 (y')^2 -10 C''y^4 (y')^2
-15 C (C'')^2y^4
\notag\\
&-12 C^3 y^2 (y'')^2+8 C y^4 (y'')^2-36 C C' C^{(3)}y^4  
+24 C^2 C^{(3)}y^3 y' -2C^{(3)}  y^5 y'  
\notag \\
&\left.
-12 C^2 C^{(4)} y^4+7 y^6 C^{(4)}
\right\}
\label{eq:drp4}
\end{align}
up to total derivative terms.
Here $y:=\sqrt{C(x)^2-G(x)}$.
We denote $p_n^{(1)}$ for the WKB coefficients of the solution to (\ref{eq:qswc_dr1}) and $p_n^{(2)}$ for (\ref{eq:qsw_dr2}).
We find that their difference arises from the pole term at $x=0$. 
For example, we find
\begin{align}
p^{(2)}_2&=p^{(1)}_2-{\partial_x G\over G}{\partial_x B\over 16 (B^2-1)^{3\over2} }.
\end{align}

We next study the quantum SW curve for the $(A_1, D_r)$-type AD theory.
After taking the scaling limit, the SW curve becomes (\ref{11.4.40}) and the 
SW differential 
is given by (\ref{eq:swdiff_dr_ad}). 
Then we quantize the curve by $\xi\rightarrow -i\hbar \pa_{\tilde{x}}$.
Since the differential equation is the same form as that of $A_r$,
the WKB solutions can be obtained as (\ref{eq:pn_ar_ad}).
We compare these quantum corrections to those obtained from the scaling limit 
of $D_r$-type gauge theory.
Taking the scaling limit (\ref{eq:scaling_dr}), the quantum correction to the classical period
scales as
\begin{align}
    p^{(1)}_{2n}dx&= \delta^{\frac{(1-2n)r}{(r-1)}} \Lambda^{1-2n}\tilde{p}_{2n}d\tilde{x}+\cdots.
    \label{eq:scaling_dr1}
\end{align}
where $\tilde{p}_{2n}$ is the WKB solution of (\ref{eq:swdiff_dr_ad}).

Now we will calculate the scaling limit of the quantum SW periods (\ref{1.16.1})-(\ref{eq:drp4}).
It turns out that the relation (\ref{eq:scaling_dr1}) does not hold for the quantum curve (\ref{eq:qsw_dr2}) due to the contribution from the pole at $x=0$.
In order to reproduce this scaling limit, we need to add the quantum correction to the potential term 
as
\begin{align}
    W(\tilde{x},\tilde{s}_i)\rightarrow W(\tilde{x},\tilde{s}_i)+\hbar^2 {A\over \tilde{x}^2},
    \label{eq:sp_dr_cor}
\end{align}
which corresponds to the shift of the parameter $\tilde{s}_r\rightarrow \tilde{s}_r+\hbar^2 A$.
One can compute the quantum correction to the WKB expansion by the replacement (\ref{eq:sp_dr_cor}) and re-expansion in $\hbar$. 
The $\tilde{p}_2$ and $\tilde{p}_4$ are modified as
\begin{align}
    \tilde{p}_2(A)&={1\over 48\sqrt{2}}{\partial_{\tilde{x}}^2W\over W^{3\over2}}+{A\over 2\sqrt{2}}{1\over \tilde{x}^2 W^{1\over2}},
\\
\tilde{p}_4(A)&=
-{7\over 3072\sqrt{2}}{(\partial_{\tilde{x}}^2W)^2\over W^{7\over2}}
+{1\over 1536\sqrt{2}} {\partial_{\tilde{x}}^4W\over W^{5\over2}}
-{A\over 64\sqrt{2}} {\partial_{\tilde{x}}^2W \over \tilde{x}^2W^{5\over2}}
+\left({(1-A)A\over 16\sqrt{2}}\right) {1\over \tilde{x}^4W^{3\over2}},
\end{align}
up to total derivatives.
Then 
 we see that in the scaling limit 
$p_{2n}^{(2)}$ scales as
\begin{align}
    p^{(2)}_{2n}dx&= \delta^{\frac{(1-2n)r}{(r-1)}} \Lambda^{1-2n}\tilde{p}_{2n}\left(A=-1/2\right)d\tilde{x}+\cdots .
    \label{eq:p2-tp}
\end{align}
Therefore we need quantum corrections to the superpotential for
the $(A_1,D_r)$-type AD theory, where the correction depends on the quantization condition of the original gauge theory before taking the limit.

We compute the differential operators which connects $\tilde{p}_{2n}(A)$
to $\tilde{p}_0$.
By the  analysis similar to the case of  $A_r$, we find that
\begin{align}
\tilde{p}_2(A)
&={1\over 24}
\left( (2r-2) (2r-3) \pa_{\ts_1}\pa_{\ts_{r-1}} 
+\sum_{i=1}^{r-1} (2r-2i-2)(2r-2i-3) \ts_i  \pa_{\ts_{i+1}}\pa_{\ts_{r-1}} 
+6 \ts_r \pa_{\ts_r}^2
\right)\tilde{p}_0
\nonumber\\
&-{A\over2} \pa_{\ts_{r}}\tilde{p}_0 .
\label{eq:o2_dr}
\end{align}
We can also write down a formula for $\tilde{p}_4(A)$, which is given in the appendix. 

For $r=3$, we find that the quantum curve with $A=0$ is equivalent to the SW curve (\ref{eq:qc_exp}) of the AD theory associated with $SU(2)$ $N_f=2$ SQCD, where
the moduli parameters are identified as
$\tilde{s}_1=-2\tilde{m}$, $\tilde{s}_2=-\tilde{u}+{4\tilde{m}^2\over3}$,
$\tilde{s}_3=-{\tilde{C}_2\over 4}$.

We next discuss the $r=4$ case. 
This AD theory is also obtained from the scaling limit of $SU(2)$ $N_f=3$ theory \cite{Argyres:1995xn,Xie:2012hs}, where the SW curve is 
\begin{align}
    & {1\over2}\left(z+{G(x)\over z}\right)=C(x), \nonumber \\
    & C(x)=-2\tilde{m}x -\tilde{u},\quad G(x)=-x^3-\tilde{C}_2 x-\tilde{C}_3,
\end{align}
where $\tilde{u}$ and $\tilde{m}$ are the operator and the coupling with the scaling dimensions being $\frac{3}{2}$ and $\frac{1}{2}$. The Casimirs of the $U(3)$ flavor symmetry $\tilde{C}_2$ and $\tilde{C}_3$ have the scaling dimensions $2$ and $3$, respectively. 
The SW differential is given by (\ref{11.3.24}).
In order to compare two curves, it is convenient to parametrize the superpotential as
\begin{align}
W(\tx,\tu_i)&=x^6-\tu_1 \tx^4 -\left( \tu_2-{\tu_1^2\over4} \right) \tx^2-\tu_3+{\tu_1\tu_2\over 6} -{\tu_4^2\over \tx^2},
\end{align}
where $\tu_i$'s are given by
\begin{align}
    \tu&=\tu_4,\ \tilde{m}=-{\tu_1\over4},\quad
    \tilde{C}_2=-4\tu_3^2+{\tu_2^2\over3},\quad
    \tilde{C}_3=-{2\tu_2^3\over 27}-{8\tu_3^2\tu_2\over3}.
\end{align}
Then with the help of the Picard-Fuchs equations satisfied by $\tilde{p}_0$, the relations (\ref{eq:o2_dr}) and (\ref{eq:o4_dr}) are shown to become
\begin{align}
\tilde{p}_2(A)&=\left(2 \pa_{\tilde{u}_1}\pa_{\tilde{u}_3}+{\tu_1^2\over 12} \pa_{\tilde{u}_3}^2 +{1+4A \over 16 \tu_4}\pa_{\tilde{u}_4}\right)
\tilde{p}_0+d(*),
\label{eq:p2a}\\
\tilde{p}_4(A)&=
\left\{
-{4\over15}\tu_1 \pa_{\tilde{u}_3}^3-{7\over 1440}\tu_1^4\pa_{\tilde{u}_3}^4-{7\over30}\tu_1^2\pa_{\tilde{u}_1}\pa_{\tilde{u}_3}^3
-{14\over5}\pa_{\tilde{u}_1}^2\pa_{\tilde{u}_3}^2
\right\}\tp_0
\nonumber\\
&-{(1+4A) \over 96 \tu_4(-\tu_2^2+12\tu_4^2)}
\left\{ 4 \tu_1^3\tu_2\tu_4 \pa_{\tilde{u}_3}^4+3 (\tu_1^2\tu_2-4\tu_2^2+48 \tu_4^2)\pa_{\tilde{u}_1}\pa_{\tilde{u}_3}\pa_{\tilde{u}_4}
\right\}\tp_0
\nonumber\\
&+{\left(4A+1\right)^2 \over 512\tu_4^3} (\pa_{\tilde{u}_4}-\tu_4\pa_{\tilde{u}_4}^2) \tp_0+d(*).
\label{eq:p4a}
\end{align}
For $A=-{1\over4}$, which is different from $-{1\over2}$ in (\ref{eq:p2-tp}), we can show that (\ref{eq:p2a}) and (\ref{eq:p4a}) correspond to those of the $SU(2)$ $N_f=3$ AD theory \cite{Ito:2018hwp}.
Therefore, for the $(A_1, D_4)$-type AD theory
obtained from the $D_4$-type gauge theory and the $SU(2)$ $N_f=3$ gauge theory,
the universality of the quantum SW curves holds if one include the quantum correction to the superpotential.

\section{Conclusions and discussion}
In this paper, we studied the quantum SW curves for the $(A_1, A_r)$ and $(A_1,D_r)$-type AD theories
and its relation to the scaling limit of the gauge theories.
We confirmed that for the $A_r$ type AD theories the quantization condition is consistent with the 
scaling limit.
We have checked the universality of the quantum SW curves of the $(A_1,A_2)$ and $(A_1,A_3)$ type AD theories.
For $D_r$-type, we found the quantum corrections to the superpotential which is obtained by the 
shift of the moduli parameter $s_r$.
By choosing the shift parameter appropriately, we found the universality of the $(A_1,D_4)$ AD theory at the quantum level.
We presented a general procedure to obtain differential operators for higher
order quantum corrections to the classical SW periods.
In order to study non-perturbative structure of the WKB expansion, 
it is necessary to explore higher order corrections explicitly\cite{Dunne:2016qix,Basar:2015xna,Basar:2017hpr,Grassi:2018bci,Ito:2017ypt,Ito:2018eon}.
It is interesting to compare the quantum periods with those obtained from the $D_r$-type Nekrasov partition function \cite{Marino:2004cn,Nekrasov:2004vw,Nakamura:2014nha}.

We can extend the present analysis to the quantum SW periods for the AD theories obtained from the scaling limit of $SU(N_c)$ SQCD \cite{IKO}.
It is interesting to check whether similar universality holds for other AD theories in the NS limit.
It is also interesting to study the case of exceptional gauge groups.
Finally, the quantum corrections to the superpotential give rise to the monodromy of the solution at the origin, which plays an important role in the study of the  ODE/IM correspondence \cite{Bazhanov:1998wj,Dorey:1999uk}.

\subsection*{Acknowledgements}
We would like to thank Hongfei Shu for useful discussion.
The work of K.I. is supported in part by Grant-in-Aid for Scientific Research 15K05043, 18K03643 and 16F16735 from Japan Society for the Promotion of Science (JSPS).

\appendix
\section*{Appendix $\tilde{p}_4(A)$ in $(A_1,D_r)$-type AD theory}
In this appendix we write down the formula for $\tilde{p}_4(A)$ in $(A_1,D_r)$-type AD theory.
\begin{align}
\tilde{p}_4(A)&=
-{7\over 5760}
 \Bigl\{
(2r-2)^2(2r-3)^2 \tilde{\pa}_1^2 \tilde{\pa}_{r-1}^2 +12 (2r-2) (2r-3) \ts_r \tilde{\pa}_1 \tilde{\pa}_{r-1} \tilde{\pa}_r^2
+36 \ts_r^2 \tilde{\pa}_r^4
\nonumber\\
&+2 (2r-2)(2r-3) \sum_{i=1}^{r-1} (2r-2i-2)(2r-2i-3) \ts_i  \tilde{\pa}_1\tilde{\pa}_{i+1} \tilde{\pa}_{r-1}^2 
\nonumber\\
&+12 \sum_{i=1}^{r-1} (2r-2i-2)(2r-2i-3) \ts_i  \ts_r\tilde{\pa}_{i+1} \tilde{\pa}_{r-1} \tilde{\pa}_r^2
\nonumber\\
&+ \sum_{i=1}^{r-1}\sum_{j=1}^{r-1} (2r-2i-2)(2r-2i-3) (2r-2j-2)(2r-2j-3) \ts_i  \ts_j
\tilde{\pa}_{i+1}\tilde{\pa}_{j+1} \tilde{\pa}_{r-1}^2 
\Bigr\} \tilde{p}_0
\nonumber\\
&-{1\over 1152}
\Bigl\{
{(2r-2)!\over (2r-6)!} \tilde{\pa}_2 \tilde{\pa}_{r-1}^2
+\sum_{i=1}^{r-3} {(2r-2i-2)!\over (2r-2i-6)!} \ts_i \tilde{\pa}_{i+2}  \tilde{\pa}_{r-1}^2
+120 \ts_r \tilde{\pa}_{r}^3
\Bigr\} \tilde{p}_0
\nonumber\\
&+{A\over 48}
\Bigl\{
(2r-2) (2r-3) \tilde{\pa}_{1}\tilde{\pa}_{r-1} \tilde{\pa}_{r}
+\sum_{i=1}^{r-1} (2r-2i-2)(2r-2i-3) \ts_i  \tilde{\pa}_{i+1}\tilde{\pa}_{
r-1} \tilde{\pa}_{r}
+6 \ts_r \tilde{\pa}_{r}^3
\Bigr\}\tilde{p}_0
\nonumber\\
&+\left({(1-A)A\over 16\cdot 2}\right) 4 \tilde{\pa}_r^2\tilde{p}_0.
\label{eq:o4_dr}
\end{align}
Here we have defined $\tpa_i=\tilde{\partial}_{\tilde{s}_i}$.



\end{document}